\documentclass[conference]{IEEEtran}
\usepackage{multirow}
\usepackage[font=scriptsize]{caption}
\usepackage{cite,amsmath,amssymb}
\usepackage{graphicx,color}
\usepackage{epstopdf}
\usepackage{multirow}
\usepackage{xspace}
\usepackage{subfigure}
\usepackage{url}
\usepackage{xspace}

\def\one{({\em i}\/)\xspace}
\def\two{({\em ii}\/)\xspace}
\def\three{({\em iii}\/)\xspace}
\def\four{({\em iv}\/)\xspace}

\def\eg{\emph{e.g.,} \xspace}
\def\etc{\emph{etc.}\xspace}
\def\ie{\emph{i.e.,}\xspace}
\def\vs{\emph{vs}\xspace}

\begin{document}

\title{Exploring Media Bias and Toxicity in South Asian Political Discourse}



\author{\IEEEauthorblockN{Adnan Qayyum\IEEEauthorrefmark{1},
Zafar Gilani\IEEEauthorrefmark{2},
Siddique Latif\IEEEauthorrefmark{1} and
Junaid Qadir\IEEEauthorrefmark{1}
}
\IEEEauthorblockA{\IEEEauthorrefmark{1}Information Technology University, Punjab, Pakistan\\ Email: (adnan.qayyum,siddique.latif,junaid.qadir)@itu.edu.pk}
\IEEEauthorblockA{\IEEEauthorrefmark{2}University of Cambridge, United Kingdom\\
Email: szuhg2@alumni.cam.ac.uk}}

\maketitle

\begin{abstract}
Media outlets and political campaigners recognise social media as a means for widely disseminating news and opinions. 
In particular, Twitter is used by political groups all over the world to spread political messages, engage their supporters, drive election campaigns, and challenge their critics.
Further, news agencies, many of which aim to give an impression of balance, are often of a particular political persuasion which is  reflected in the content they produce. Driven by the potential for political and media organisations to influence public opinion, our aim is to quantify the nature of political discourse by these organisations through their use of social media.
In this study, we analyse the sentiments, toxicity, and bias exhibited by the most prominent Pakistani and Indian political parties and media houses, and the pattern by which these political parties utilise Twitter.
We found that media bias and toxicity exist in the political discourse of these two developing nations.  

\end{abstract}


\section{Introduction}
\label{sec:introduction}

Twitter has assumed the role of an influential micro-blogging platform worldwide with 328 million active users per month, as of December 2017.
One of the major manifestations of this influence can be seen in the socio-political realm in which social media outlets like Twitter play a major role in shaping the views, opinions, and sentiments of people \cite{eltantawy2011arab}.
This `socio-political' influence---driven by the wide reach of social media platforms, and the propensity for content to 'go viral'--- has had a significant impact on the interplay among the society, mainstream political powerhouses, and the mainstream media.

While the developed world faces issues of political influence through social media \cite{udanor2016determining}.
`Fake news', `post-truth' and `bots' have become a topic of serious discussion in the Western world, where judiciary, law and order, democracy and free speech are fundamental.
However, \emph{concerns over the use of social media for political aims are compounded in the developing world}, where the democratic apparatus is comparatively weak coupled with the power of the emotionally charged, politically motivated mass crowds \cite{eltantawy2011arab,lynch2014syria}.


{\bf Goals of this research:} Twitter is a popular social media service that is widely being used as a source of online news. The open nature and global reach of such news media can spread the fake or misleading, and agenda-driven information to deceive people and influence their opinions \cite{zannettou2017web}. It has been argued that the existence of bias in social media towards or against a certain entity or ideology can shape the behaviour of a certain group of people, and also inadvertently assist to certain agenda to be fulfilled \cite{Zafar:2016:SAC:3001913.3006644}. Such bias can also affect the voting behaviour and have the ability to promote intolerance and antagonisms in social and political issues \cite{saez2013social,dellavigna2007fox,chiang2011media,groseclose2005measure}. The awareness, tracking, and overcoming bias in news media is very crucial for societies especially in emerging countries, where social networking services like Twitter has the capability to shape the democracy \cite{leichty2016twitter,6758880}. 

Therefore in this paper, we focus on two South Asian developing economies for Twitter news media analysis using sentiments, bias, and toxicity in the political landscape.
We consider three pressing questions: 
\begin{enumerate}
    \item Does the media's reporting\slash commentary through social media, as regards to political entities, indicate bias? And if so, how and to what degree?
    \item If provocative\slash inflammatory \emph{(toxic)} speech is a common practice on social media in developing political economies?
    \item To which extent these political economies exhibit similarity in social media usage? 
\end{enumerate}

We choose two biggest countries in South Asia: Pakistan and India, with a combined population of 1.517 billion people as of 2016. These countries share high geographical as well as cultural and political similarity as two different nations. As Pakistan and India come into existence as two separate states by the division of colonial India. This makes these countries an interesting choice for exploring similar trends in their political sphere with the fact that social media has had a direct impact on their political landscapes. 
For instance, social media, particularly Twitter have contributed to the PTI's (political party) success to emerge as a third largest national party in the 2013 Pakistan general elections \cite{6758880}. While the AAP's (political party in India) emergence as a political party was also driven by social media \cite{leichty2016twitter}. Indeed, we see the use of social media by political parties is on the rise in both nations. However, we only use Twitter accounts that are officially associated with news organisations (news channels) and political parties for our analysis. 


{\bf Contributions of this paper:} In this paper, we build upon our previous work on measuring sentiments of controversial topics \cite{Zafar:2016:SAC:3001913.3006644} to explore the South Asian political landscape.
Through a detailed empirical analysis, 
we outline a methodology to 
understand and measure: \one~the coverage and sentiments biases among news organisations for political parties, \two~the intensity of toxicity in news coverage given by news organisations to political parties, and also by political parties towards their rivals, \three~the common topics discussed by political parties, and \four~comparison of similarity trends of Twitter usage among Pakistani and Indian politicians.

{\bf Terminology and definitions:} We first define the terms 
sentiment, bias, and toxicity.\\
\textit{Bias} is an inclination or prejudice toward a person or a group \cite{bias-def}. This might be for or against, but considered to be unfair either way. In particular, we study two types of bias, \ie coverage bias, and statement bias.    
We define coverage bias as the preference given to one news over the expense of another and statement bias as the coverage sentiment given by a news channel to a political entity, \ie reporting using favourable terms and phrases \cite{d2000media}.\\
\textit{Sentiment} is a view, feeling or an expression that a person holds with regard to some aspects of a topic \cite{hovy2015sentiment}.
This could both be positive (optimistic) or negative (pessimistic), or utopian \vs dystopian.\\
\textit{Toxicity} is the attribute of something being rude, disrespectful, malignant, hateful, \etc \\

{\bf Paper arrangement:} The rest of the paper is organised as follows. Section \ref{sec:relatedWork} presents the related work.
Details about the dataset and methodology adapted for analysis is presented in Section \ref{sec:methodology}. 
Empirical analysis with results is presented in Section \ref{sec:analysis_results} and we conclude in Section \ref{sec:conclusion}.

\section{Related Work}
\label{sec:relatedWork}
In recent years many researchers have  analysed Twitter data for prediction of political events such as electoral prediction \cite{burnap2016140}, political orientation prediction \cite{conover2011predicting} and bias of news sources \cite{saez2013social}.
Despite having high popularity among its viewership, mainstream media often fails to meet the standards of journalism ethics, and inadvertently patronises certain ideas \cite{Zafar:2016:SAC:3001913.3006644}. 
Furthermore, news sources are also known to inject political bias while reporting \cite{younus2012investigating,an2012visualizing, groseclose2005measure}, potentially affecting the political beliefs of the audience, which might result in altering voting behaviour \cite{an2012visualizing}. Recently it was shown that Fox News had been misrepresenting facts in an effort to appeal to conservative viewers \cite{saez2013social}. 

Here we present the literature related to media biases quantification and the work done using social media on the politics of Pakistan and India.

\textit{Media Bias:} In \cite{saez2013social}, authors introduced unsupervised methods to quantify three types of media biases: selection bias, coverage bias, and statement bias. They analysed the data from different international news channels for the duration of two weeks and showed that the biases are observable and depend upon the geographical boundaries. Similarly, numerous works have quantified media bias but most of them are focused on US-centric news outlets \cite{an2012visualizing,groseclose2005measure}. This US-centric research is restricted to measure media bias towards two political ideologies (\ie liberal and conservative) which cannot be applied for measuring media bias in more diverse multi-party settings like in Pakistan and India. 
Therefore, in this research, we analyse the nature of political discourse among political 
organisations of Pakistan and India in addition to only measure media biases. 

\begin{table*}
\centering
\scriptsize
\caption{Tweets collected per entity, $P$ is a political party, $NC$ is a news channel and \#Accs denote number of accounts, collected over a period of six months.}

\begin{tabular}{|l|l|l|r|r|r|r|r|r|r|r|r|}
\hline
{\bf Country}                                        & {\bf Type}       & {\bf Entity} & {\bf \#Accs} & {\bf \#Tweets} & \multicolumn{2}{c|}{\bf \#English} & \multicolumn{2}{c|}{\bf \#Urdu/Hindi} & {\bf \#Processed} & {\bf Klout} & {\bf \#Followers} \\ \hline
\multicolumn{1}{|l|}{\multirow{8}{*}{Pakistan}} & \multirow{4}{*}{P}  & Pakistan Tehreek-e-Insaf (PTI) & 9 & 22113 & 17126 & 77.45\% & 3718 & 16.81\% & 20844 & 68.22 & 2281771 \\ \cline{3-12} 
\multicolumn{1}{|l|}{}  & & Pakistan Muslim League Nawaz (PML-N) & 10 & 15516 & 8928	& 57.54\% & 4736 & 30.52\% & 13664 & 65.4 & 1276272 \\ \cline{3-12}
\multicolumn{1}{|l|}{}  & & Pakistan People Party (PPP) &  12 & 6831 & 4124	& 60.37\% & 1592 & 23.31\% & 5716 & 65.33 & 762349 \\ \cline{3-12} 
\multicolumn{1}{|l|}{}  &  & Muttahida Qaumi Movement (MQM) & 8 & 1522 & 972	& 63.86\% & 406 & 26.68\% & 1378 & 54.75 & 85006 \\ \cline{2-12} 
\multicolumn{1}{|l|}{}                         & \multirow{4}{*}{NC} &   Geo News & 9 & 8648 & 5417 & 62.64\% & 2830 & 32.72\% & 8247 & 67.4 & 1506823 \\ \cline{3-12} 
\multicolumn{1}{|l|}{} & & ARY News & 8 & 6914 & 3287 & 47.54\% & 2514 &  36.36\% & 5801 & 67.12 & 929566 \\ \cline{3-12} 
\multicolumn{1}{|l|}{} & & Express News & 8 & 4397 & 2872	& 65.32\% & 1208 & 27.47\% & 4080 & 59.22 & 409202\\ \cline{3-12} 
\multicolumn{1}{|l|}{}  & & DAWN News & 8 & 3346 & 2481 & 74.15\% & 559 & 16.71\% & 3040 & 61 & 394896 \\ \hline
\multirow{8}{*}{India}                            & \multirow{4}{*}{P}  & Bharatiya Janata Party (BJP) & 13 & 103167 & 75165 & 72.86\% & 19084 & 18.5\% & 94249 & 75 & 6390782 \\ \cline{3-12} 
                                               &                     & Indian National Congress (INC) & 12 & 54590 & 40027 & 73.32\% & 12260 & 22.46\% & 52287 & 71.42 & 813516 \\ \cline{3-12} 
                                               &                     & All India Trinamool Congress (AITC) & 5 & 3280 & 2740 & 83.54\% & 186 & 5.67\% & 2926 & 53.4 &  403568 \\ \cline{3-12} 
                                               &                     & Aam Aadmi Party (AAP) & 11 & 21411 & 9344 & 43.64\% & 10511	& 49.09\% & 19855 & 65.73 & 403568 \\ \cline{2-12} 
                                               & \multirow{4}{*}{NC} & Aaj Tak News & 10 & 27126 & 8608 & 31.73\% & 16855	& 62.14\% & 25463 & 66 & 1320962 \\ \cline{3-12} 
                                               &                     & NDTV News & 7 & 40866 & 28530 & 69.81\% & 5176	& 12.67\% & 33706 & 66 & 1764582 \\ \cline{3-12} 
                                               &                     & Zee News & 9 & 12778 & 4786 & 37.46\% & 6972 & 54.56\% & 11758 & 62 & 663970 \\ \cline{3-12} 
                                               &                     & India Today News & 8 & 26251 & 22650 & 86.28\% & 1846 & 7.03\% & 24496 & 72.4 & 2379074 \\ \hline

\end{tabular}
\label{tab:dataset}
\end{table*}

\textit{Pakistani Political Landscape and Social Media:} The study presented in \cite{6758880}, explores the effective use of social media by a Pakistani political party. 
The authors specifically focus on the Pakistan Tehreek-e-Insaaf's (PTI) campaigns during the 2013 general elections and showed how the PTI 
emerged as a third national party by mobilising and engaging voters on social media. 
The efficacy of inferring political behaviour from the Twitter analysis is studied in \cite{razzaq2014prediction}. 
Large-scale experimentation on sentiment analysis and tweet classification showed that the political behaviour of party followers and campaigning impact can be predicted using social media content. In a similar study \cite{younus2014election}, authors proposed a methodology to extract sentiment from the political conversation using Twitter data. They performed sentiment analysis on political slang words and political trolls that supporters of different political parties used to attack each other during 2013 general elections of Pakistan. 


\textit{Indian Political Landscape and Social Media:} Various studies aim at examining social media content to explore political landscape of India, especially in the era of Indian general elections. 
In \cite{jaidka20152014}, the effect of changing political traditions is investigated before 2014 Indian general election by using the tweets from the official accounts of top ten political parties for a period of two months. The study found that, among ten parties, Bharatiya Janta Party (BJP), and Aam Aadmi Party (AAP) were actively using Twitter for campaigning and projecting their party manifesto. Such extensive use of Twitter by BJP's politicians aided them to win a majority in Lok Sabha of India.  
Sentiment analysis of tweets for prediction of 2016 general state elections is presented in \cite{sharma2016prediction}. Authors measured sentiments of the population towards five political parties by analysing data of two months. A comprehensive analysis of 2014 Indian elections is presented in \cite{lu2014india} that specifically focuses on BJP success which was greatly influenced by the effective use of Twitter. 

This work differs from the studies presented above which mostly aim at exploring Pakistani and Indian political landscape for electoral events.  To the best of our knowledge, this is first attempt to explore political inclination of media in these countries and no previous work directly compares politicians and their Twitter usage pattern across these nations. 



\section{Data and Methodology}
\label{sec:methodology}

In this section, we describe our data collection process and provide details about the dataset. 
Moreover, methodology adapted to empirically answer the questions raised earlier using different approaches is presented in this section. 

\subsection{Data collection and data pre-processing}
\label{subsec:dataset}

\textbf{Data curation:} Our dataset consists of tweets curated from accounts of news channels and political parties using Stweeler~\cite{gilani2016stweeler}. Data was curated for six months from June to November 2017. For this, we identified four major political parties of both countries using statistics of last general elections of Pakistan\footnote{https://goo.gl/9iPAgK} and India\footnote{https://goo.gl/dZbMo3} held in 2013 and 2014 respectively. We then identify popular members of each political party including parties' chairmen, secretaries, and members of parliament and national assembly. Similarly, four major news channels from both countries are selected based on the higher number of viewers. We consider the journalists as the representatives of their respective news channel. Because they are known to reflect political inclination \cite{bhuwan-political-inclination,donsbach2004political} of their employer \cite{zaller1999theory}. 

We considered only those accounts who have official accounts verified by Twitter (with the exception of a few) and have high influence. 
One way is to select influential account based on a large number of followers (as adopted in \cite{ali2018measuring}), but past research has shown that a high number of followers on Twitter does not always entail an influential account \cite{cha2010measuring}. Therefore, we used well-known Klout score \cite{rao2015klout} to calculate the influence of Twitter users. It is an influence scoring system that is used to assigns scores to millions of users across different social media networks on a daily basis \cite{rao2015klout}. It assigns a score to a user, ranging from 1 to 100, where a higher score means more influence. To determine the influence of an entity having multiple accounts, we obtain an {\em entity Klout score} by averaging Klout score of all accounts belonging to this entity. 

Based on the statistics, we replaced third ranked party, \ie All India Anna Dramuka Kazhagam (AIADMK) of 2014 India's general election with Aam Aadmi Party (AAP), as AAP has a strong influence on Twitter as compared to AIADMK, \ie AIADMK has only three official twitter accounts with comparatively little followers base and have less Klout score.
This decision can be attributed by a recent study \cite{leichty2016twitter} which aims at investigating how AAP emerged as a national party from anti-corruption social movement and became a popular political party. 


\textbf{Data pre-processing:}
After downloading the tweets corpus, we performed pre-processing of the data to standardise it for effective textual analysis. 
It involves removing all ancillary elements such as stop words, punctuation signs, unnecessary spaces, URL, user mention symbol, hashtag symbol and converting upper-case letters to lower case letters.


Note that our dataset also contains tweets in languages other than English, most prominently Urdu and Hindi, which are the national languages of Pakistan and India respectively. The language in these tweets was first detected using Google's language detection library\footnote{https://github.com/Mimino666/langdetect} and then translated into English using widely used (e.g., \cite{manushree2017comparative, iguider2017language} ) TextBlob's\footnote{https://goo.gl/yvMPLn} language translation library.
Also, by following this strategy tweets in languages other than English, Urdu and Hindi are filtered out which reduced the total size of the dataset, as shown in Table \ref{tab:dataset}.  
To avoid redundancy, we only considered tweets having unique content after performing pre-processing and translation.

\textbf{Data statistics:} Details about the dataset such as size, number of accounts, average Klout score and the average number of followers for each media and political entity of both countries are shown in Table \ref{tab:dataset}.  
We see that among all the Pakistani political parties, PTI is the most influential political party on Twitter, while MQM is the least influential. 
Similarly, Geo News is the most influential news channel on Twitter, while Dawn News is the least one. On the other hand, BJP is the most influential political party in India and AITC is less influential. Among Indian news channels, India Today News has higher influence as compared to others and Zee News has the least influence. From Table \ref{tab:dataset}, it is also clear that PTI and BJP are the frequent tweeter and have large follower base as compared to the other parties in Pakistan and India respectively. There is a significant difference in the number of tweets by PTI and BJP, as BJP's members are more active on social media.




\subsection{Methodology}
\label{subsec:method}
We leveraged from three types of experiments, \ie \one measuring bias of news channels, \two toxicity quantification, and \three measuring similarities in social media usage by political parties. 
Details about each of the analysis are given as follows.

\subsubsection{Measuring bias of news channels}
\label{subsubsec:bias}
To measure the degree of `bias' of news channels towards political parties,  we examine two types of biases, \ie coverage bias and statement bias on news channels' data. 

\paragraph{Measuring coverage bias} The degree of policies to cover different political issues or stories among different news sources may vary. Coverage bias comes in when a specific news organisation gives more prominence or attention to a certain political party as compared to other news media. The distribution of prominence given to particular political entity or party can be quantified in various ways. We compute such coverage bias by measuring the instances of members' mentions of each Pakistani and Indian political party in tweets corpus collected from respective national news channels.  Mentions are calculated only for those members who are part of this study, \ie from whom we collect data and we use members' names and their Twitter account's screen name for mentions measurement. 
Coverage computation is further enhanced by measuring instances of party names' acronym and a few important key phrases, \eg pm abbasi, cm kpk and cm punjab, \etc 
In coverage analysis, a tweet is marked relevant for a political party if it either mentions its members, name acronym, or a key phrase.

\paragraph{Measuring statement bias} Statement bias exists in tweets when more favourable statements are used for a particular political party or entity at the expense of others. We study statement biases in tweets identified relevant in the coverage bias analysis by using sentiments. We computed sentiments score in three dimensions  (i.e., positive, negative and neutral) using TextBlob's sentiment analyser. It assigns a sentiment score $s_t$ ranging from $-1.0$ to $1.0$, where a value close to these numbers means to contain more negative and positive sentiment respectively. 
Let's assume $s_t$ is the sentiment score for a tweet $t$ and to classify $s_t$ into any of three sentiment types, \ie \textit{positive, negative, and neutral}, we choose interval given in Eq. \ref{eq:1} where, $S_t$ is the classified sentiment type. 

\begin{equation}\label{eq:1}
    S_t = \bigg\{ \begin{tabular}{ccc}
                 $s_t$ $>$= 0.1 & positive \\
                 $s_t$ $<$= -0.1 & negative \\
                 otherwise & neutral
                \end{tabular}
\end{equation}

Although a rare case, if a tweet contains mentions related to two or more entities then that tweet will be marked relevant for each entity.
The tweet will then be classified based on its sentiment. However, if it was only positively addressed towards one of the entities (and negatively towards the others), sentiment scores of other entities would also be positively affected. 
We opt to ignore this limitation for now, since the overall percentage of tweets affected by this is negligible, \ie only 0.03\% of 116591 tweets are affected in our dataset.


\subsubsection{Quantifying toxicity} 
\label{subsubsec:toxicity}
The toxic instances like verbal violence and aggression in political tweets widely used due to the increase of political polarisation. To quantify toxicity of tweets, we used Perspective API\footnote{https://www.perspectiveapi.com/}. Perspective API is a part of Google's Conversation AI project which enables the developers to identify the 'toxicity' of the given social media content. We indicated the toxicity of tweets having negatively reporting by news channels about political parties of both countries.
Also we quantify the intensity of toxicity of the tweets by different Pakistani and Indian parties when talking about their respective rival political parties. 

We note that toxicity is an indication; the Perspective API has some limitations as it is designed for text written in US English style and it fails to compute toxicity for text having bad sentence structure. Therefore, tweets for which API was successfully able to compute toxicity are incorporated into the analysis.

\subsubsection{Measuring similarity in social media usage}

To explore and measure the similarity trend exhibited by political entities of Pakistan and India in their social media usage, we perform political discourse analysis. As the definition of politics is ambiguous hence it is hard to define political discourse as well. Therefore, we focus our analysis on discovering major things or issues discussed by each political party of both countries by using topic modelling approach. 
We leveraged from widely used topic modelling technique, \ie Latent Dirichlet Allocation (LDA) \cite{blei2003latent}. It is an unsupervised generative probabilistic model which tends to discover latent structure in a set of documents by representing each document as a random collection of latent topics. Where each topic itself is modelled as a distribution over words co-occurrences. We performed topic modelling using the aggregated data from the accounts of each political party, as they are its representatives and will often talk about their parties' manifesto and news. The data is then transformed into term frequencies for LDA model training. We used grid search cross-validation in order to tune model's hyperparameters and for selecting the best-estimated model for topic modelling of our data. As a result, we got the best-trained model with two topics for each political entity.  
In addition to LDA, we also used principal component analysis (PCA) to measure the similarity in Twitter usage by political entities of Pakistan and India.



Note that ethical considerations such as the four principles as proposed in the Belmont Report \cite{us1979belmont} and the Menlo Report \cite{dittrich2012menlo} and described by Salganik namely \cite{salganik2017bit}, \one respect for persons \two beneficence \three justice and \four respect for law and public interest were taken into account during this work.
Although informed consent was not taken from individuals whose tweets are being used, as we utilised public data and mitigated informational risk, \ie the potential harm from the disclosure of personal information through aggregation of data where appropriate.


\begin{figure*}[!ht]
\centering
\subfigure[]{\includegraphics[trim=3cm 0.0cm 3cm 1cm,clip=true,width=9cm]{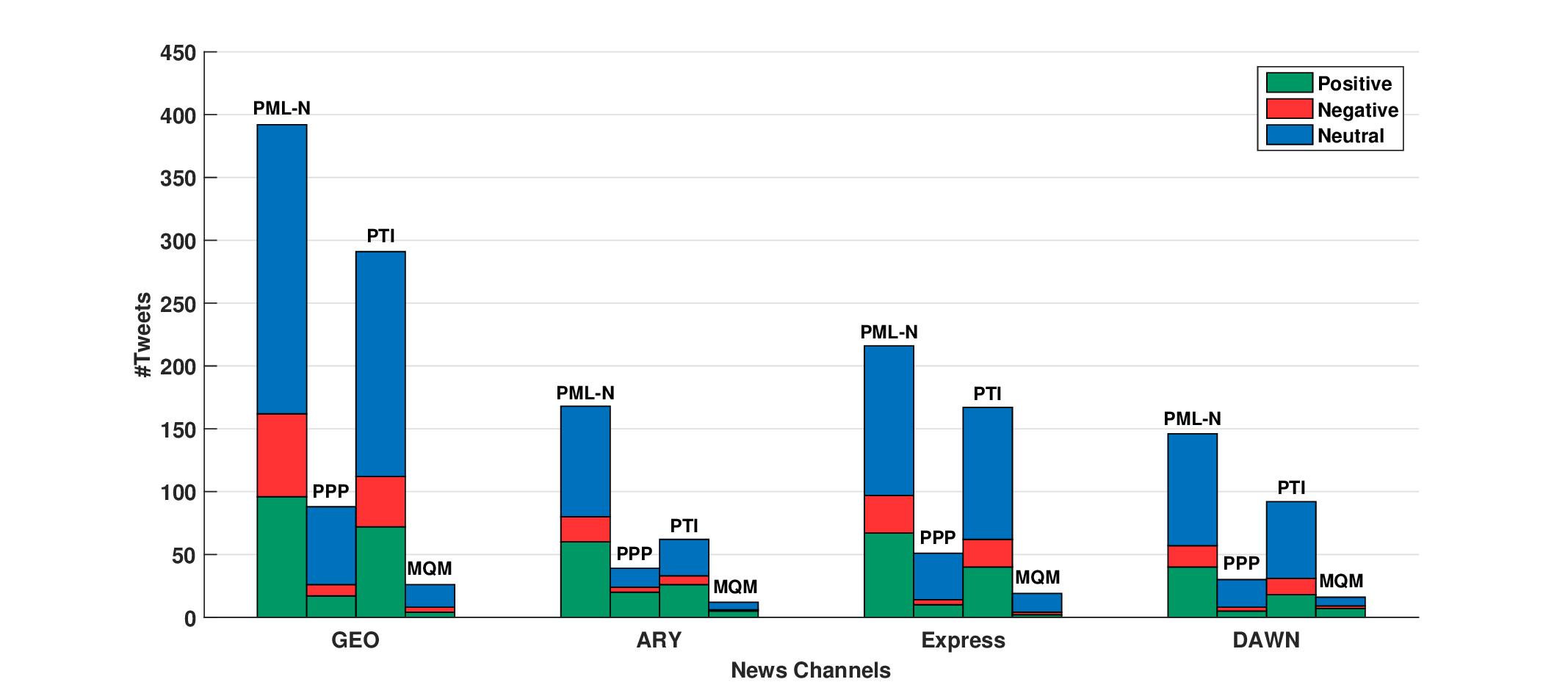}\label{fig:pk_coverage}}
\subfigure[]{\includegraphics[trim=2.8cm 0.0cm 3cm 1cm,clip=true,width=9cm]{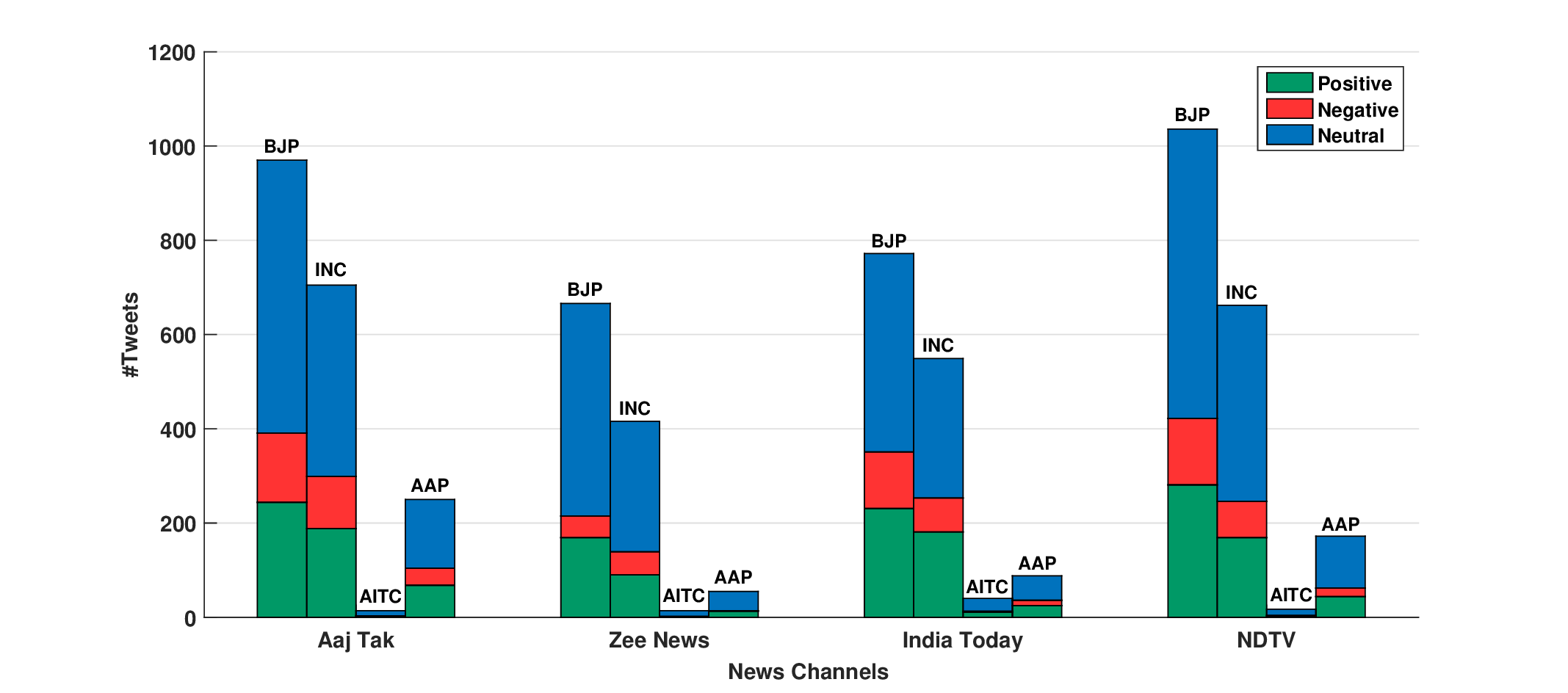}\label{fig:in_coverage}}
\caption{Coverage and sentiments given by news channels to political parties. Where Figure \ref{fig:pk_coverage} shows the coverage and sentiments given by various Pakistani news channels to their national political parties and \ref{fig:in_coverage} present the coverage and sentiments given by Indian news channels to their national political parties.}
\label{fig:pk_in_coverage}
\end{figure*}

\section{Analysis and Results}
\label{sec:analysis_results}
In this section, detailed empirical analysis for exploring bias, sentiments, toxicity, and political discourse along with results is presented.

\subsection{Indicating political bias in online news sources}
\label{subsec:bias}

Here we discuss the coverage bias and sentiments of coverage given to each political party by each news channel.

\textbf{Which political party is given the most and the least coverage by each news channel?}

Figure \ref{fig:pk_coverage} shows the coverage given to different Pakistani political parties by Pakistani news channels. 
It is prominent from Figure \ref{fig:pk_coverage} that almost all news channels give most attention to PML-N which is expected as it holds the government in federal and in a major province. 
Whereas, PTI is the second party to get more prominence and it is clear that all news channels give less coverage to PPP and very least coverage to MQM. 
We hypothesise that PML-N, PTI, and PPP are given more media attention compared to MQM due to the fact that all hold the major governmental unit and would understandably get more coverage: PML-N (federal govt and Punjab provincial), PTI (KPK provincial), and PPP (Sindh provincial).

Similarly, the coverage given to various Indian political parties by Indian news channels is shown in Figure \ref{fig:in_coverage}.
It is clear from the Figure \ref{fig:in_coverage} that, every channel gives most coverage to BJP while, INC gets more air as compared to AITC, and AAP but less than BJP. AITC is getting very less coverage, and Zee News give the least coverage to AAP when compared with other channels. 
BJP's and INC's higher coverage may be due to the fact that, BJP is ruling Indian Government while INC is the second largest party of Indian parliament and it has ruled government in the previous decade.

\begin{table*}[!ht]
\centering
\scriptsize
\captionsetup{width=.75\textwidth}
\caption{Average toxicity given by Pakistani and Indian news channels to their national political parties. Where, \%Neg shows percentage of negative tweets, Avg Tox means average toxicity and \%NP represents percentage of tweets that is not processed.}
\begin{tabular}{|l|l|l|l|l|l|l|l|l|l|l|l|l|}
\hline
\multirow{3}{*}{\textbf{\begin{tabular}[c]{@{}l@{}}Pakistani\\ Parties\end{tabular}}} & \multicolumn{12}{c|}{\textbf{Pakistani News Channels}}                                                                                                                                                                           \\ \cline{2-13} 
                                                                                      & \multicolumn{3}{c|}{\textbf{GEO}}                                                                                & \multicolumn{3}{c|}{\textbf{ARY}}                                                                                & \multicolumn{3}{c|}{\textbf{Express}}                                                                            & \multicolumn{3}{c|}{\textbf{DAWN}}                                                                               \\ \cline{2-13} 
                                                                                      & \multicolumn{1}{c|}{\textbf{\%Neg}} & \multicolumn{1}{c|}{\textbf{Avg Tox}} & \multicolumn{1}{c|}{\textbf{\%NP}} & \multicolumn{1}{c|}{\textbf{\%Neg}} & \multicolumn{1}{c|}{\textbf{Avg Tox}} & \multicolumn{1}{c|}{\textbf{\%NP}} & \multicolumn{1}{c|}{\textbf{\%Neg}} & \multicolumn{1}{c|}{\textbf{Avg Tox}} & \multicolumn{1}{c|}{\textbf{\%NP}} & \multicolumn{1}{c|}{\textbf{\%Neg}} & \multicolumn{1}{c|}{\textbf{Avg Tox}} & \multicolumn{1}{c|}{\textbf{\%NP}} \\ \hline
\textbf{PML-N}                                                                        & 16.84                               & 0.2783                                & 0                                  & 11.9                                & 0.2917                                & 0                                  & 13.89                               & 0.2847                                & 0                                  & 11.64                               & 0.3061                                & 0                                  \\ \hline
\textbf{PPP}                                                                          & 10.23                               & 0.323                                 & 0                                  & 10.26                               & 0.4325                                & 0                                  & 7.84                                & 0.502                                 & 0                                  & 10                                  & 0.3118                                & 0                                  \\ \hline
\textbf{PTI}                                                                          & 13.75                               & 0.3101                                & 0                                  & 11.29                               & 0.3793                                & 0                                  & 13.17                               & 0.3317                                & 0                                  & 14.13                               & 0.2525                                & 0                                  \\ \hline
\textbf{MQM}                                                                          & 15.38                               & 0.1744                                & 0                                  & 8.33                                & 0.3965                                & 0                                  & 10.53                               & 0.4249                                & 0                                  & 12.5                                & 0.1544                                & 0                                  \\ \hline
\multirow{3}{*}{\textbf{\begin{tabular}[c]{@{}l@{}}Indian\\ Parties\end{tabular}}}    & \multicolumn{12}{c|}{\textbf{Indian News Channels}}                                                                                                                                                                                                                                                                                                                                                                                                                       \\ \cline{2-13} 
                                                                                      & \multicolumn{3}{c|}{\textbf{Aaj Tak}}                                                                            & \multicolumn{3}{c|}{\textbf{Zee News}}                                                                           & \multicolumn{3}{c|}{\textbf{India Today}}                                                                        & \multicolumn{3}{c|}{\textbf{NDTV}}                                                                               \\ \cline{2-13} 
                                                                                      & \multicolumn{1}{c|}{\textbf{\%Neg}} & \multicolumn{1}{c|}{\textbf{Avg Tox}} & \multicolumn{1}{c|}{\textbf{\%NP}} & \multicolumn{1}{c|}{\textbf{\%Neg}} & \multicolumn{1}{c|}{\textbf{Avg Tox}} & \multicolumn{1}{c|}{\textbf{\%NP}} & \multicolumn{1}{c|}{\textbf{\%Neg}} & \multicolumn{1}{c|}{\textbf{Avg Tox}} & \multicolumn{1}{c|}{\textbf{\%NP}} & \multicolumn{1}{c|}{\textbf{\%Neg}} & \multicolumn{1}{c|}{\textbf{Avg Tox}} & \multicolumn{1}{c|}{\textbf{\%NP}} \\ \hline
\textbf{BJP}                                                                          & 15.15                               & 0.2784                                & 2.16                               & 6.91                                & 0.3353                                & 0.75                               & 15.54                               & 0.3595                                & 0.78                               & 13.61                               & 0.3504                                & 0.97                               \\ \hline
\textbf{INC}                                                                          & 15.74                               & 0.2472                                & 0.99                               & 11.78                               & 0.2849                                & 0.24                               & 13.11                               & 0.3282                                & 0.36                               & 11.63                               & 0.3104                                & 1.21                               \\ \hline
\textbf{AITC}                                                                         & 0                                   & 0                                     & 0                                  & 7.14                                & 0.4065                                & 0                                  & 5                                   & 0.318                                 & 0                                  & 11.76                               & 0.2489                                & 0                                  \\ \hline
\textbf{AAP}                                                                          & 14.4                                & 0.266                                 & 0.8                                & 1.82                                & 0.2352                                & 0                                  & 12.5                                & 0.3553                                & 0                                  & 10.47                               & 0.2964                                & 0.58                               \\ \hline
\end{tabular}
  
  \label{tab:toxicity}
\end{table*}
\begin{table*}[!ht]
\centering
\scriptsize
\captionsetup{width=.75\textwidth}
\caption{Average toxicity given by each political party to their rivals. Where, \%Tweets is the percentage of relevant toxic tweets, Avg Tox means average toxicity and \%NP represents percentage of tweets that is not processed.}
\label{tab:par_par}
\begin{tabular}{|l|c|l|l|l|l|l|l|l|l|l|l|l|}
\hline
\multirow{3}{*}{\textbf{\begin{tabular}[c]{@{}l@{}}Pakistani \\ Rival\\ Parties\end{tabular}}} & \multicolumn{12}{c|}{\textbf{Pakistani Political Parties}}\\\cline{2-13} 
& \multicolumn{3}{c|}{\textbf{PML-N}} & \multicolumn{3}{c|}{\textbf{PPP}} & \multicolumn{3}{c|}{\textbf{PTI}} & \multicolumn{3}{c|}{\textbf{MQM}}          \\ \cline{2-13} 
                                                                                               & \textbf{\%Tweets}            & \multicolumn{1}{c|}{\textbf{Avg Tox}} & \multicolumn{1}{c|}{\textbf{\%NP}} & \multicolumn{1}{c|}{\textbf{\%Tweets}} & \multicolumn{1}{c|}{\textbf{Avg Tox}} & \multicolumn{1}{c|}{\textbf{\%NP}} & \multicolumn{1}{c|}{\textbf{\%Tweets}} & \multicolumn{1}{c|}{\textbf{Avg Tox}} & \multicolumn{1}{c|}{\textbf{\%NP}} & \multicolumn{1}{c|}{\textbf{\%Tweets}} & \multicolumn{1}{c|}{\textbf{Avg Tox}} & \multicolumn{1}{c|}{\textbf{\%NP}} \\ \hline
\textbf{PML-N}                                                                                 & \multicolumn{3}{c|}{-}                                                                                 & 5.35                                & 0.2497                                & 0.04                               & 13.4                                & 0.2681                                & 0.04                               & 4.26                                & 0.1854                                & 0                                  \\ \hline
\textbf{PPP}                                                                                   & \multicolumn{1}{l|}{0.9}  & 0.3274                                & 0.02                               & \multicolumn{3}{c|}{-}                                                                                           & 1.6                                 & 0.2466                                & 0                                  & 3.39                                & 0.1953                                & 0.11                               \\ \hline
\textbf{PTI}                                                                                   & \multicolumn{1}{l|}{4.38} & 0.2739                                & 0.07                               & 4.08                                & 0.2738                                & 0                                  & \multicolumn{3}{c|}{-}                                                                                           & 5.36                                & 0.1964                                & 0                                  \\ \hline
\textbf{MQM}                                                                                   & \multicolumn{1}{l|}{0.08} & 0.231                                 & 0                                  & 1.03                                & 0.2243                                & 0.12                               & 0.25                                & 0.2393                                & 0                                  & \multicolumn{3}{c|}{-}                                                                                           \\ \hline
\multirow{3}{*}{\textbf{\begin{tabular}[c]{@{}l@{}}Indian\\ Rival\\ Parties\end{tabular}}}     & \multicolumn{12}{c|}{\textbf{Indian Political Parties}}                                                                                                                                                                                                                                                                                                                                                                                                         \\ \cline{2-13} 
                                                                                               & \multicolumn{3}{c|}{\textbf{BJP}}                                                                      & \multicolumn{3}{c|}{\textbf{INC}}                                                                                & \multicolumn{3}{c|}{\textbf{AITC}}                                                                               & \multicolumn{3}{c|}{\textbf{AAP}}                                                                                \\ \cline{2-13} 
                                                                                               & \textbf{\%Tweets}            & \multicolumn{1}{c|}{\textbf{Avg Tox}} & \multicolumn{1}{c|}{\textbf{\%NP}} & \multicolumn{1}{c|}{\textbf{\%Tweets}} & \multicolumn{1}{c|}{\textbf{Avg Tox}} & \multicolumn{1}{c|}{\textbf{\%NP}} & \multicolumn{1}{c|}{\textbf{\%Tweets}} & \multicolumn{1}{c|}{\textbf{Avg Tox}} & \multicolumn{1}{c|}{\textbf{\%NP}} & \multicolumn{1}{c|}{\textbf{\%Tweets}} & \multicolumn{1}{c|}{\textbf{Avg Tox}} & \multicolumn{1}{c|}{\textbf{\%NP}} \\ \hline
\textbf{BJP}                                                                                   & \multicolumn{3}{c|}{-}                                                                                 & 13.45                               & 0.266                                 & 0.66                               & 3.66                                & 0.2931                                & 0.33                               & 9.96                                & 0.2599                                & 0.47                               \\ \hline
\textbf{INC}                                                                                   & \multicolumn{1}{l|}{5.54} & 0.2679                                & 0.45                               & \multicolumn{3}{c|}{-}                                                                                           & 1.77                                & 0.3102                                & 0.2                                & 2.94                                & 0.2629                                & 0.21                               \\ \hline
\textbf{AITC}                                                                                  & \multicolumn{1}{l|}{0.11} & 0.3498                                & 0.01                               & 0.11                                & 0.3421                                & 0.01                               & \multicolumn{3}{c|}{-}                                                                                           & 0.08                                & 0.3067                                & 0.02                               \\ \hline
\textbf{AAP}                                                                                   & \multicolumn{1}{l|}{0.8}  & 0.3027                                & 0.11                               & 0.47                                & 0.3822                                & 0.08                               & 0.98                                & 0.2822                                & 0.26                               & \multicolumn{3}{c|}{-}                                                                                           \\ \hline
\end{tabular}
\end{table*}

\textbf{What are sentiments of coverage given to each political party by different news channels?} 

Figure \ref{fig:pk_in_coverage} also shows the sentiments (i.e., positive, negative or neutral) given by various Pakistani and Indian news channels when covering each national political party. 
Almost every Pakistani news channel is giving most positive coverage to PML-N. Whereas, PTI is the second party to get higher positive coverage and consequently PPP is at number three after PML-N and PTI.  Also, MQM is having very least positive and most neutral coverage by almost every channel. What is surprising is ARY's more positive coverage towards PML-N. The common perception in Pakistan is that GEO News is consistently pro-PML-N, and anti-PTI, while ARY News is pro-PTI and anti-PML-N. This is despite the fact that PML-N has had a fair share of political controversies as well as corruption scandals, including the latest Panama Papers\footnote{\url{https://en.wikipedia.org/wiki/Panama_Papers_case_(Pakistan)}}. Also, prime minister of PML-N stepped down by Supreme Court's disqualification order on account of his corruption scandals.

Similarly, the coverage sentiments given by various Indian news channels when reporting each Indian political party is illustrated in Figure \ref{fig:pk_in_coverage}. It is found that every Indian news channel is giving more positive sentiments to BJP.  While INC is getting more positive coverage as compared to AITC and AAP but lesser then BJP. BJP gets more negative coverage from almost every channel with exception of Zee News.  Also, AITC is the only party to get a fewer coverage and most neutral coverage among others.

\subsection{Exploring toxicity in social discussions}
\label{subsec:toxicity}

To indicate provocative, \ie toxic speech in social discussions by news channels and political parties we consider two questions.  

\textbf{How toxic is the content when a news channel negatively discusses different political parties?}

To answer this question, we compute average toxicity of tweets 
having negative sentiment given by various news channels of Pakistan and India to political parties. 
Results for toxicity analysis on content dissemination about each political party by different news channels of both countries are summarised in Table \ref{tab:toxicity}. It reveals that Pakistani channels such as GEO News are spreading relatively more toxic content about PTI, ARY News and Express News are reporting PPP with higher toxicity despite their more negative coverage to PML-N. DAWN News is giving highest toxic sentiments to PPP. GEO's higher toxic coverage to PTI and ARY's higher negative coverage 
to PML-N support the perception as stated earlier, \ie GEO News is consistently pro-PML-N, and anti-PTI, while ARY News is pro-PTI and anti-PML-N.

Similarly, among Indian news channels, Aaj Tak News is disseminating higher toxic content about BJP and Zee News is giving higher negative coverage to INC while spreading more toxic content about AITC.  
Whereas, India Today News and NDTV are giving coverage with high toxic intensity to BJP as compared to other news channels.


\textbf{How toxic is the content when a political party talks about their other rival political parties?}

To get relevant tweets posted by a specific party about its rival parties, we measure occurrences of rival parties' members mentions using their names and Twitter screen names. This is further supplemented by measuring mentions of rival parties 'name acronym and a few important key phrases, \eg pm abbasi, cm punjab, \etc This strategy is similar to what we followed for coverage analysis of political parties by news channels.
The results of this analysis, \ie the percentage of relevant toxic tweets, average toxicity, and percentage of tweets not included 
are presented in Table \ref{tab:par_par}.
It is found that among Pakistani political parties, PML-N is talking more about PTI with comparatively high toxicity, and PPP has posted more tweets about PML-N and spreading higher toxicity about PTI similar to MQM. Among all parties, PTI is the only party which is talking more about PML-N with high toxicity. 
The common trend in Pakistan is that PTI is consistently criticising PML-N and PTI's high toxic intensity to PML-N validates this trend. 
We see that almost all parties are frequently talking about PML-N as it holds federal unit. MQM is a small party having least representatives, therefore, getting very less consideration by all other parties. In Pakistan, the PTI and PPP are major representatives of the opposition and hence they talk more about PML-N that is ruling current government of the country. 

Similarly, among Indian political parties, BJP is getting a higher number of tweets from each party with a comparatively high intensity of toxicity (see Table \ref{tab:par_par}). This is due to fact that BJP is currently ruling the government of India and it receives criticism in terms of toxicity from other parties. INC is giving high toxicity to AAP and AITC while it talks more about BJP with a comparative high toxicity. 
Similarly, AITC is spreading more toxic content about BJP and also spreading higher toxicity about INC just like the trend found in statistics of AAP about INC.


\subsection{Twitter political discourse analysis}
\label{subsec:politicaldiscourseanalysis}
Here we highlight the major things or issues discussed during data collection period by each political party on their Twitter accounts.

\textbf{What are the major issues discussed by the political parties on Twitter?} 

\begin{figure*}[t]
\centering
\subfigure[PML-N]{\includegraphics[width=0.24\linewidth]{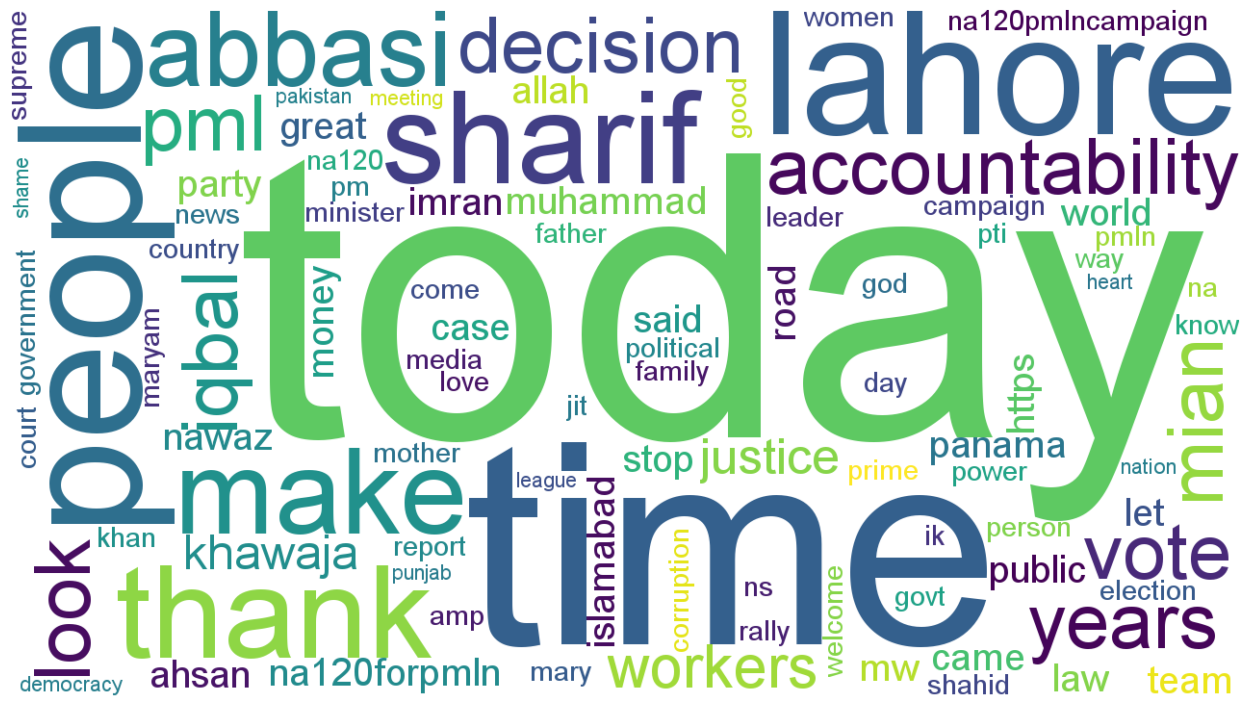}}
\subfigure[PPP]{\includegraphics[width=0.24\linewidth]{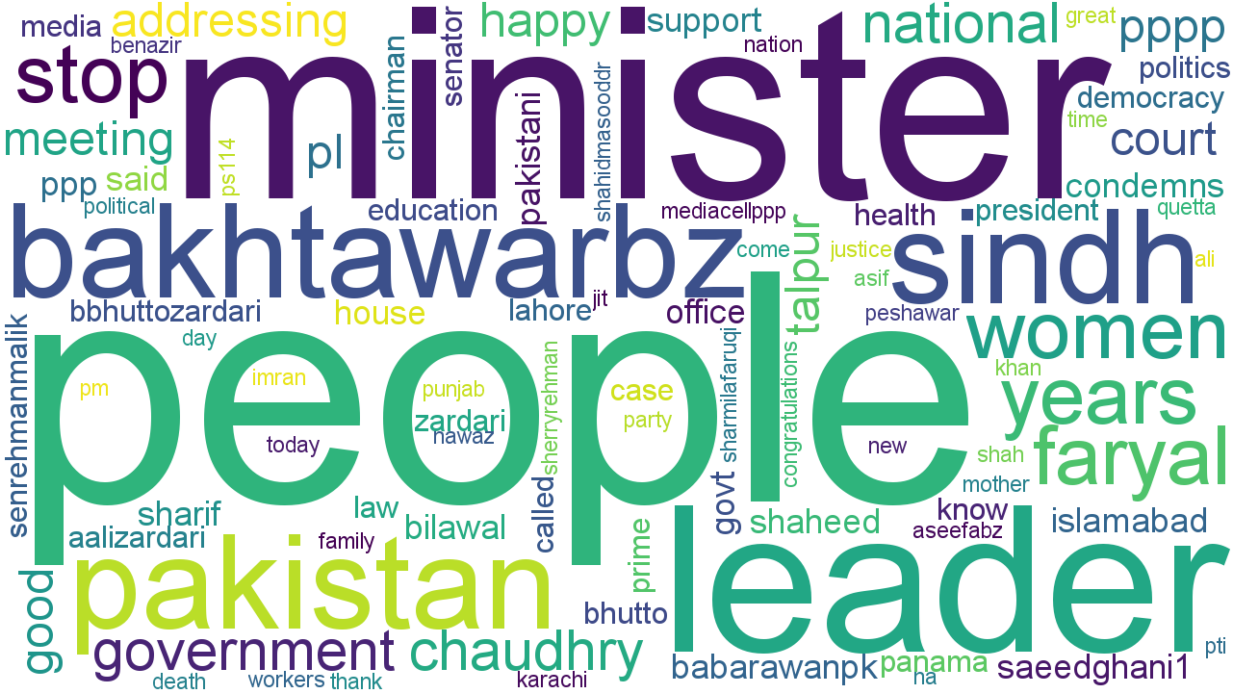}}
\subfigure[PTI]{\includegraphics[width=0.24\linewidth]{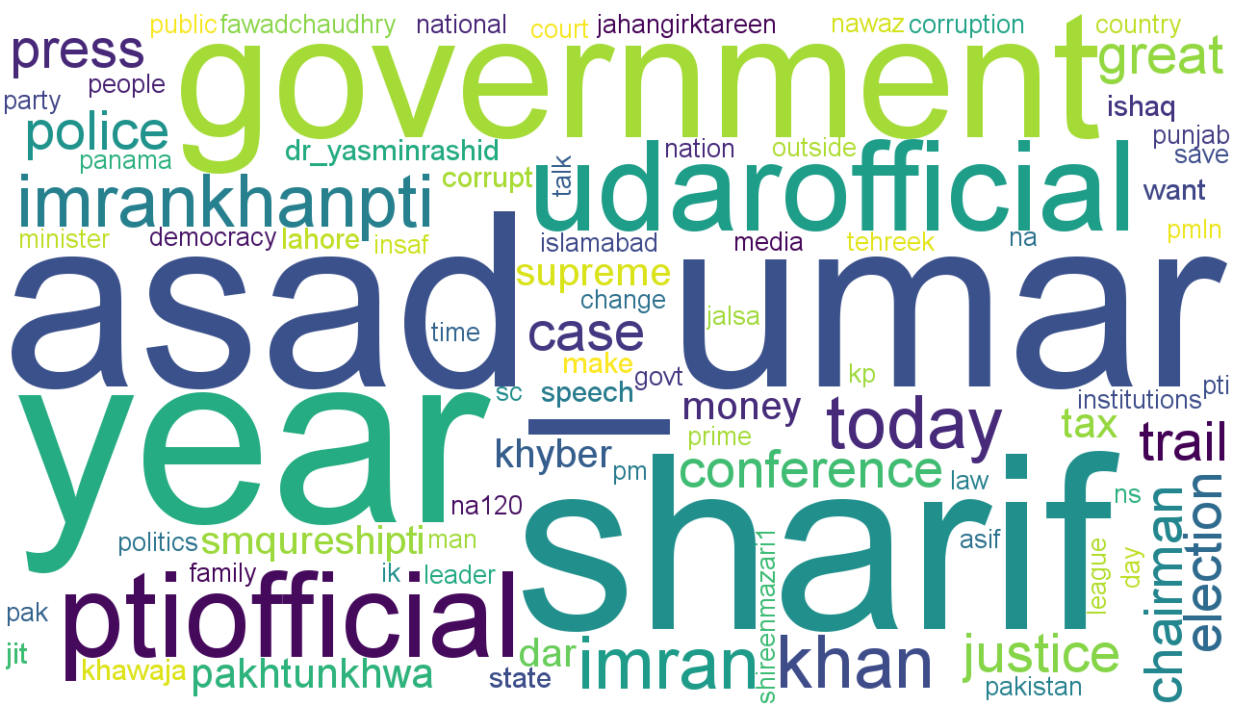}}
\subfigure[MQM] {\includegraphics[width=0.24\linewidth]{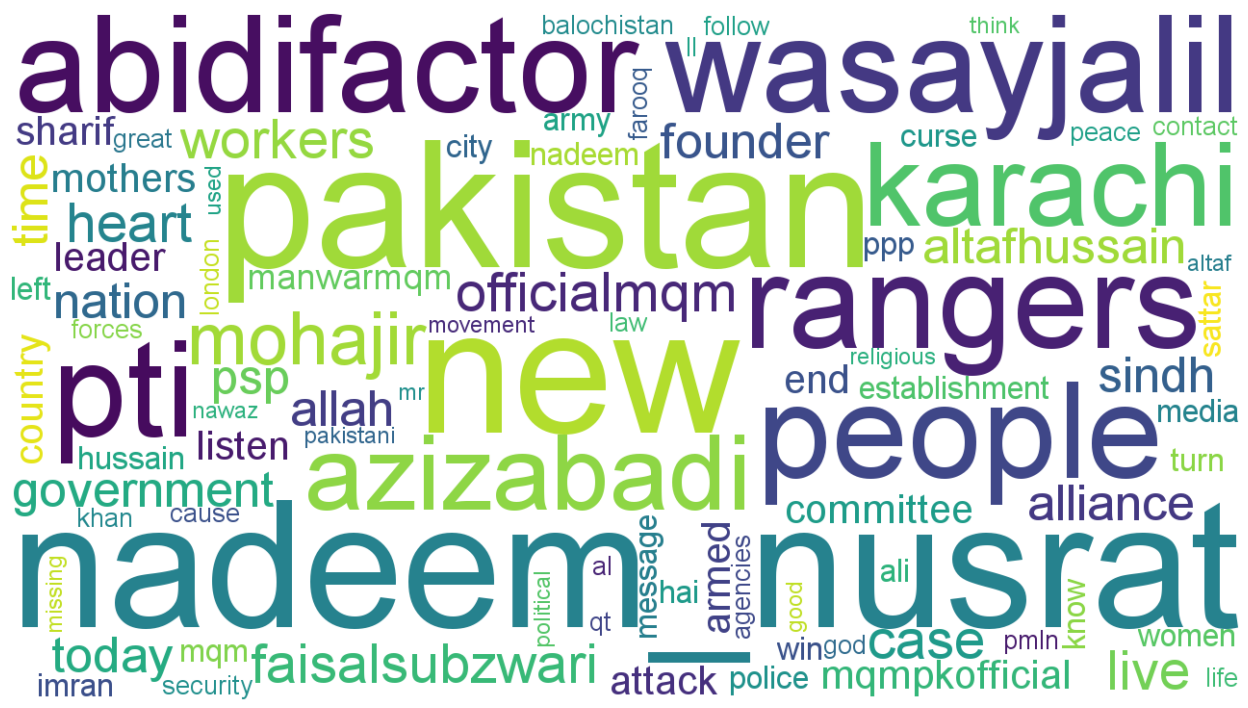}}
\subfigure[BJP]{\includegraphics[width=0.24\linewidth]{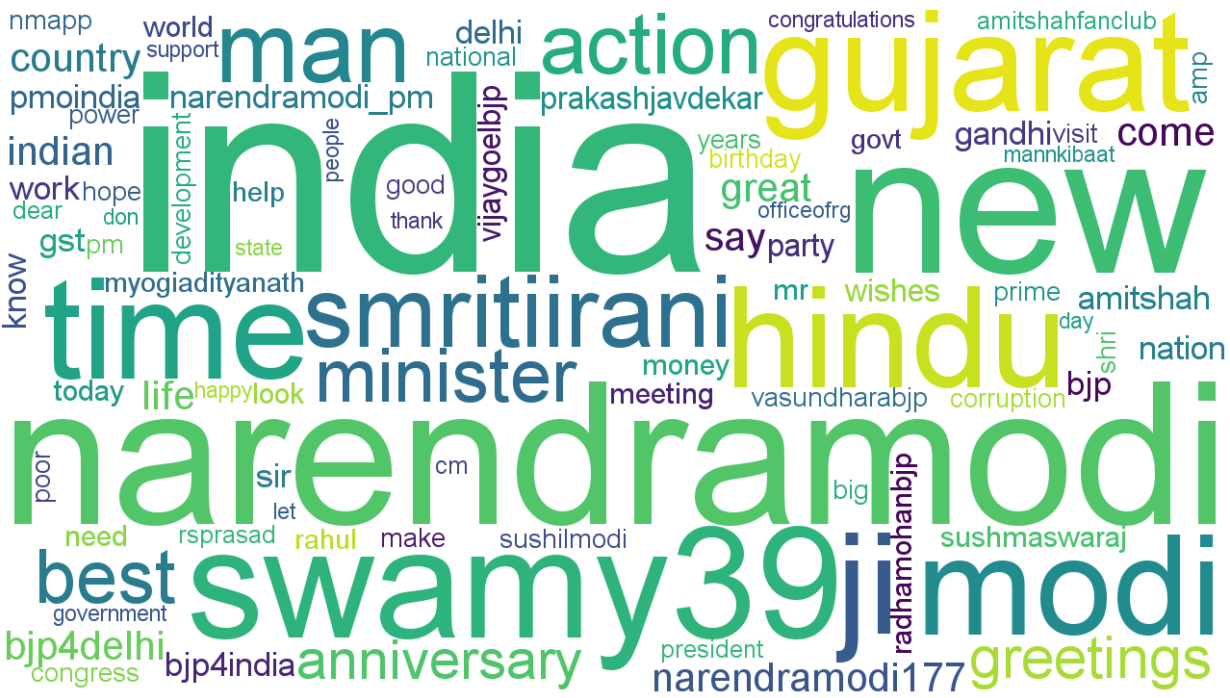}}
\subfigure[INC]{\includegraphics[width=0.24\linewidth]{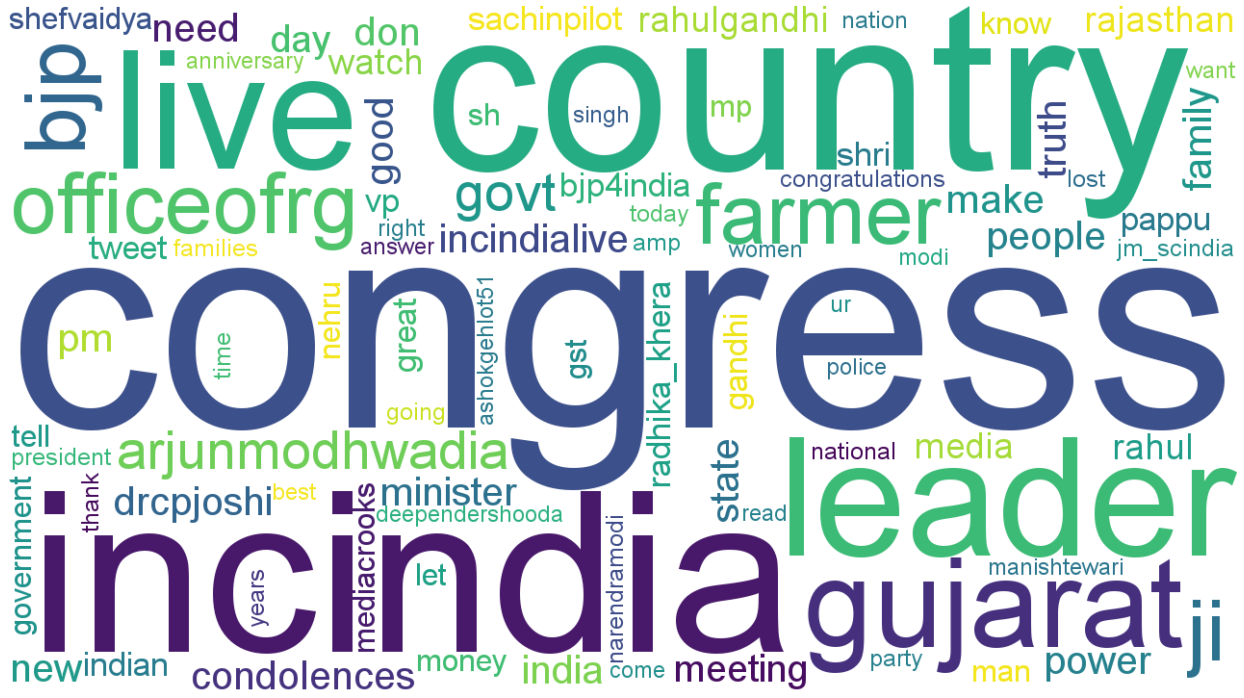}}
\subfigure[AITC]{\includegraphics[width=0.24\linewidth]{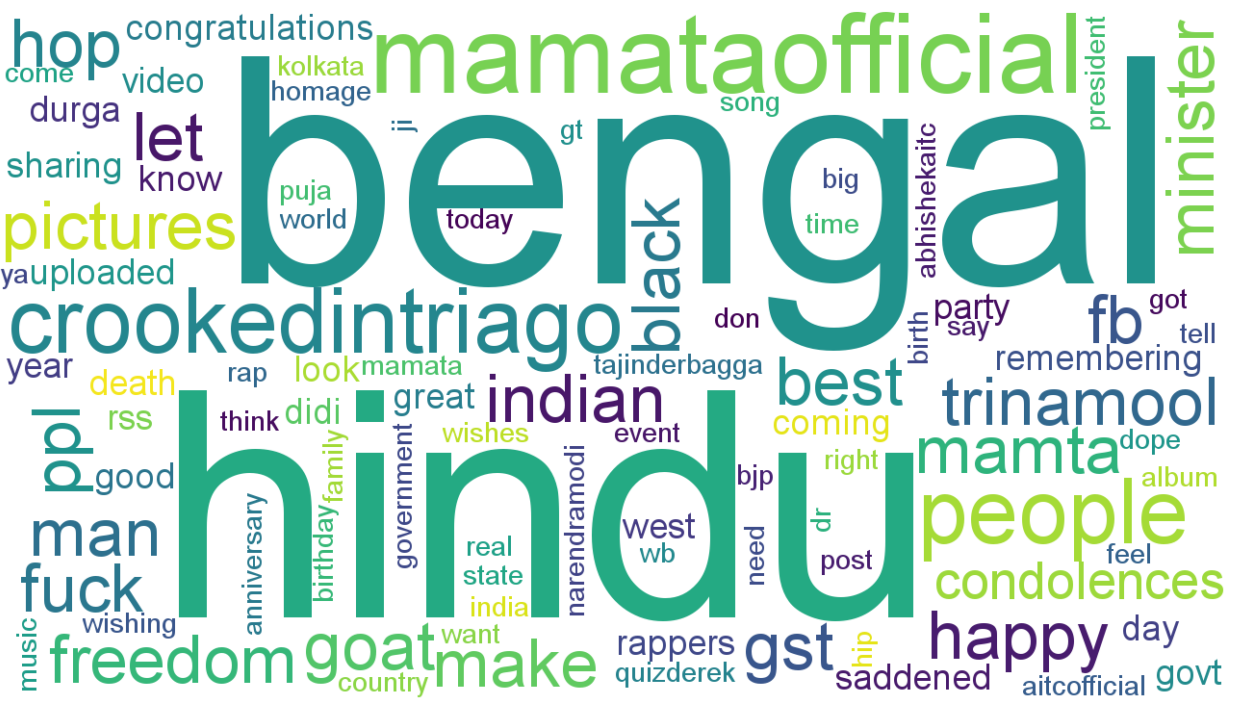}}
\subfigure[AAP]{\includegraphics[width=0.24\linewidth]{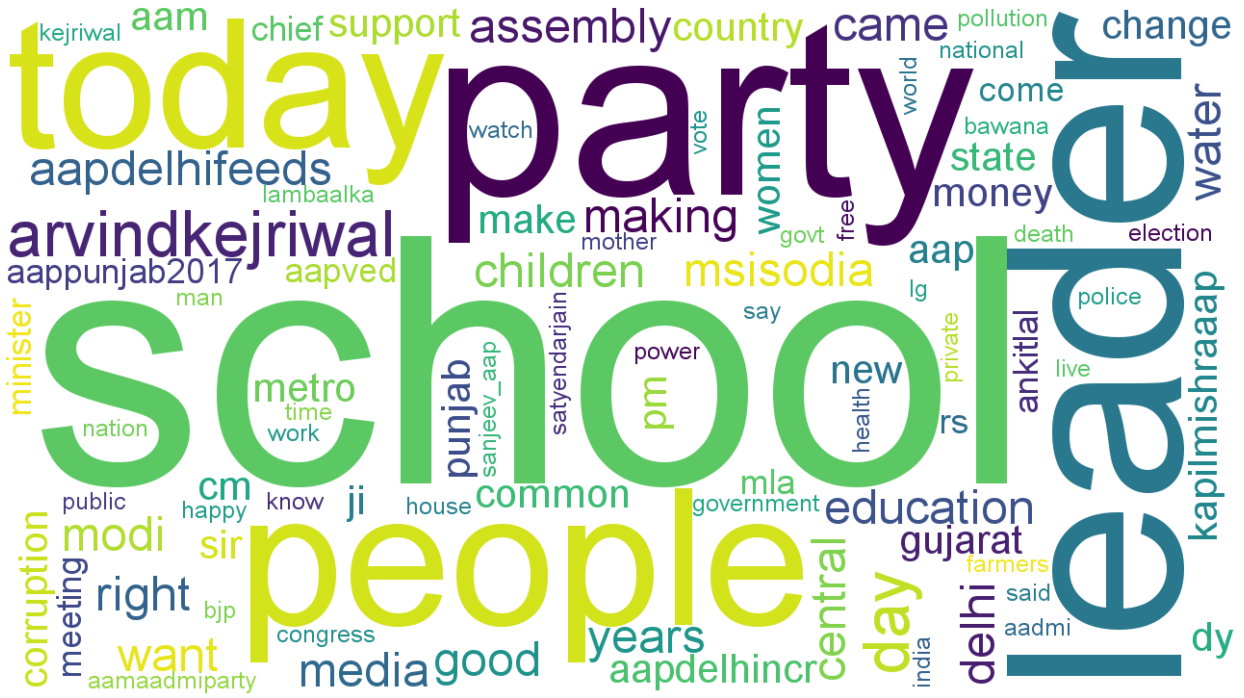}}
\caption{Word clouds illustrating common topics discussed by various  Pakistani (upper row) and Indian (lower row) political parties on Twitter.}
\label{fig:wordclouds}
\end{figure*}

To observe topics under discussion, we generate word clouds from top 50 terms associated with each topic learned by LDA model. Figure \ref{fig:wordclouds} depicts the similarities between the way incumbents (present government) and the oppositions use Twitter for political discourse. Where terms in larger size represent the high frequency of use and importance. We found that each Pakistani and Indian political party typically discusses topics such as Pakistan, India, themselves, their leadership, their members, and their opposing political parties, \etc Issues of national interests are rarely found in their conversations like they can be seen in social media usage by American political entities \cite{le2017revisiting}. 

\textbf{To which extent political entities of Pakistan and India share similarity while using Twitter sphere?}

\begin{figure}[!ht]
\centering
\includegraphics[trim=3.2cm 0.1cm 3cm 0.4cm,clip=true,width=1\linewidth]{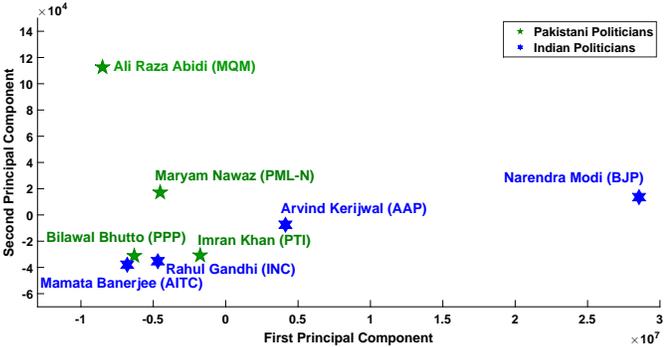}
\caption{Visualisation of similarity among various politicians of Pakistan and India using dimensionality reduction of different features using PCA.}
\label{fig:similarity}
\end{figure}

For this typical analysis, we selected only popular members of each political entity of both countries based on the number of their followers.
To measure the degree of similarity in the Twitter user profile and usage pattern, we extracted twelve user profile related features such as the number of tweets posted, number of friends and followers, \etc and few features from tweets produced by these members and their tweeting pattern, \eg number of used hashtags, and URLs \etc 
After feature extraction, we use principal component analysis (PCA) to visualise similarity exhibited by various political entities of Pakistan and India in two-dimensional space (see Figure \ref{fig:similarity}).

We discovered that among the selected politicians of Pakistan and India, Bilawal Bhutto (PPP), Imran Khan (PTI), Rahul Gandhi (INC) and Mamata Banerjee (AITC) share resemblance in their Twitter profile and usage. They are making a cluster in the left bottom of the Figure \ref{fig:similarity} with least variance (relatively high correlation) in feature space. While Maryam Nawaz (PML-N), Arvind Kerijwal (AAP), Ali Raza Abidi (MQM) and Narendra Modi (BJP) exhibit dissimilarity when compared with rest of politicians. We see that Narendra Modi's dissimilarity is due to his huge follower base, \ie more than 39 million, while Ali Raza Abidi has posted a significantly large number of Tweets which becomes a discriminative feature for him. 
\section{Conclusions and Future Work}
\label{sec:conclusion}

We studied the media biases and toxicity among different online news organisations of Pakistan and India. Further, the similarity in political discussions of these two South Asian countries is highlighted to find similar trends in Twitter usage. Our results based on the empirical analysis support the following high-level conclusions.

Political parties of both developing countries are found to be very active on social media. Our findings revealed that PTI and BJP are the dominant among all Pakistani and Indian political parties on Twitter, respectively. Among news organisations of these countries, Geo News and India Today are the dominant news channels on Twitter.

News organisations of Pakistan and India show the political leaning towards political parties in terms of coverage and statement biases. We are able to empirically observe the toxicity in tweets of the news organisation for some political parties and also in tweets of each political party towards its rivals. Interestingly, political parties holding the federal government in Pakistan and India receive more criticism and toxicity on Twitter.

The political discourse analysis shows that the political parties from both countries usually talk about the things involving the names of their countries, leadership, party members, and their rivals with a few discussions on national issues. A similarity trend in Twitter usage is also observed when popular politicians from both countries are compared. Most of the popular political entities have similarity in Twitter usage but only a few politicians are found dissimilar based on the significant difference in their follower base and frequency of tweeting.  

Future work includes increasing the size of the dataset by including more loosely associated user accounts, \ie including those entities that are less prominent but perhaps more `extreme' or confrontational in their views (as is common to the region).
Moreover, it is worth acknowledging there are several considerations in using social media to analyse the actual underlying political landscape. 
For example, social media may be a poor proxy due to the statistical selection bias, \ie the Twitter sample users may not be the true representative of the overall population. Moreover, our analysis leveraged existing tools for sentiment analysis and measuring toxicity, which were developed in an English-speaking context.
There is clear scope to explore the building and application of such tools that are tailored to the language and cultural aspects of the region being analysed.

\textbf{Acknowledgments:}
This work is partially funded by the UK EPSRC Global Challenges Research Fund (Grant EP/R512783/1).
We would like to thank Sarim Zafar and Usman Sarwar for assisting us in preparing an earlier version of the manuscript.

\bibliographystyle{IEEEtran}
\bibliography{mediabias}

\begin{thebibliography}{10}
\providecommand{\url}[1]{#1}
\csname url@samestyle\endcsname
\providecommand{\newblock}{\relax}
\providecommand{\bibinfo}[2]{#2}
\providecommand{\BIBentrySTDinterwordspacing}{\spaceskip=0pt\relax}
\providecommand{\BIBentryALTinterwordstretchfactor}{4}
\providecommand{\BIBentryALTinterwordspacing}{\spaceskip=\fontdimen2\font plus
\BIBentryALTinterwordstretchfactor\fontdimen3\font minus
  \fontdimen4\font\relax}
\providecommand{\BIBforeignlanguage}[2]{{%
\expandafter\ifx\csname l@#1\endcsname\relax
\typeout{** WARNING: IEEEtran.bst: No hyphenation pattern has been}%
\typeout{** loaded for the language `#1'. Using the pattern for}%
\typeout{** the default language instead.}%
\else
\language=\csname l@#1\endcsname
\fi
#2}}
\providecommand{\BIBdecl}{\relax}
\BIBdecl

\bibitem{eltantawy2011arab}
N.~Eltantawy and J.~B. Wiest, ``The arab spring| social media in the egyptian
  revolution: reconsidering resource mobilization theory,'' \emph{International
  Journal of Communication}, vol.~5, p.~18, 2011.

\bibitem{udanor2016determining}
C.~Udanor, S.~Aneke, and B.~O. Ogbuokiri, ``Determining social media impact on
  the politics of developing countries using social network analytics,''
  \emph{Program}, vol.~50, no.~4, pp. 481--507, 2016.

\bibitem{lynch2014syria}
M.~Lynch, D.~Freelon, and S.~Aday, ``Syria’s socially mediated civil war,''
  \emph{United States Institute of Peace}, vol.~91, no.~1, pp. 1--35, 2014.

\bibitem{zannettou2017web}
S.~Zannettou, T.~Caulfield, E.~De~Cristofaro, N.~Kourtelris, I.~Leontiadis,
  M.~Sirivianos, G.~Stringhini, and J.~Blackburn, ``The web centipede:
  understanding how web communities influence each other through the lens of
  mainstream and alternative news sources,'' in \emph{Proceedings of the 2017
  Internet Measurement Conference}.\hskip 1em plus 0.5em minus 0.4em\relax ACM,
  2017, pp. 405--417.

\bibitem{Zafar:2016:SAC:3001913.3006644}
\BIBentryALTinterwordspacing
S.~Zafar, U.~Sarwar, Z.~Gilani, and J.~Qadir, ``Sentiment analysis of
  controversial topics on pakistan's twitter user-base,'' in \emph{Proceedings
  of the 7th Annual Symposium on Computing for Development}, ser. ACM DEV
  '16.\hskip 1em plus 0.5em minus 0.4em\relax New York, NY, USA: ACM, 2016, pp.
  35:1--35:4. [Online]. Available:
  \url{http://doi.acm.org/10.1145/3001913.3006644}
\BIBentrySTDinterwordspacing

\bibitem{saez2013social}
D.~Saez-Trumper, C.~Castillo, and M.~Lalmas, ``Social media news communities:
  gatekeeping, coverage, and statement bias,'' in \emph{Proceedings of the 22nd
  ACM international conference on Conference on information \& knowledge
  management}.\hskip 1em plus 0.5em minus 0.4em\relax ACM, 2013, pp.
  1679--1684.

\bibitem{dellavigna2007fox}
S.~DellaVigna and E.~Kaplan, ``The fox news effect: Media bias and voting,''
  \emph{The Quarterly Journal of Economics}, vol. 122, no.~3, pp. 1187--1234,
  2007.

\bibitem{chiang2011media}
C.-F. Chiang and B.~Knight, ``Media bias and influence: Evidence from newspaper
  endorsements,'' \emph{The Review of Economic Studies}, vol.~78, no.~3, pp.
  795--820, 2011.

\bibitem{groseclose2005measure}
T.~Groseclose and J.~Milyo, ``A measure of media bias,'' \emph{The Quarterly
  Journal of Economics}, pp. 1191--1237, 2005.

\bibitem{leichty2016twitter}
G.~B. LEICHTY, M.~U. D'SILVA, and M.~R. JOHNS, ``Twitter and aam aadmi party:
  Collective representations of a social movement turned political party.''
  \emph{Intercultural Communication Studies}, vol.~25, no.~2, 2016.

\bibitem{6758880}
S.~Ahmed and M.~M. Skoric, ``My name is khan: The use of twitter in the
  campaign for 2013 pakistan general election,'' in \emph{2014 47th Hawaii
  International Conference on System Sciences}, Jan 2014, pp. 2242--2251.

\bibitem{bias-def}
\BIBentryALTinterwordspacing
P.~Today, ``Bias access date: 31 january 2018.'' [Online]. Available:
  \url{https://www.psychologytoday.com/basics/bias}
\BIBentrySTDinterwordspacing

\bibitem{d2000media}
D.~D'Alessio and M.~Allen, ``Media bias in presidential elections: a
  meta-analysis,'' \emph{Journal of communication}, vol.~50, no.~4, pp.
  133--156, 2000.

\bibitem{hovy2015sentiment}
E.~H. Hovy, ``What are sentiment, affect, and emotion? applying the methodology
  of michael zock to sentiment analysis,'' in \emph{Language production,
  cognition, and the Lexicon}.\hskip 1em plus 0.5em minus 0.4em\relax Springer,
  2015, pp. 13--24.

\bibitem{burnap2016140}
P.~Burnap, R.~Gibson, L.~Sloan, R.~Southern, and M.~Williams, ``140 characters
  to victory?: Using twitter to predict the uk 2015 general election,''
  \emph{Electoral Studies}, vol.~41, pp. 230--233, 2016.

\bibitem{conover2011predicting}
M.~D. Conover, B.~Gon{\c{c}}alves, J.~Ratkiewicz, A.~Flammini, and F.~Menczer,
  ``Predicting the political alignment of twitter users,'' in \emph{Privacy,
  Security, Risk and Trust (PASSAT) and 2011 IEEE Third Inernational Conference
  on Social Computing (SocialCom), 2011 IEEE Third International Conference
  on}.\hskip 1em plus 0.5em minus 0.4em\relax IEEE, 2011, pp. 192--199.

\bibitem{younus2012investigating}
A.~Younus, M.~A. Qureshi, S.~K. Kingrani, M.~Saeed, N.~Touheed, C.~O'Riordan,
  and P.~Gabriella, ``Investigating bias in traditional media through social
  media,'' in \emph{Proceedings of the 21st International Conference on World
  Wide Web}.\hskip 1em plus 0.5em minus 0.4em\relax ACM, 2012, pp. 643--644.

\bibitem{an2012visualizing}
J.~An, M.~Cha, K.~P. Gummadi, J.~Crowcroft, and D.~Quercia, ``Visualizing media
  bias through twitter,'' in \emph{Proc. ICWSM SocMedNews Workshop}, 2012.

\bibitem{razzaq2014prediction}
M.~A. Razzaq, A.~M. Qamar, and H.~S.~M. Bilal, ``Prediction and analysis of
  pakistan election 2013 based on sentiment analysis,'' in \emph{Advances in
  Social Networks Analysis and Mining (ASONAM), 2014 IEEE/ACM International
  Conference on}.\hskip 1em plus 0.5em minus 0.4em\relax IEEE, 2014, pp.
  700--703.

\bibitem{younus2014election}
A.~Younus, M.~A. Qureshi, M.~Saeed, N.~Touheed, C.~O'Riordan, and G.~Pasi,
  ``Election trolling: analyzing sentiment in tweets during pakistan elections
  2013,'' in \emph{Proceedings of the 23rd International Conference on World
  Wide Web}.\hskip 1em plus 0.5em minus 0.4em\relax ACM, 2014, pp. 411--412.

\bibitem{jaidka20152014}
K.~Jaidka and S.~Ahmed, ``The 2014 indian general election on twitter: An
  analysis of changing political traditions,'' in \emph{Proceedings of the
  Seventh International Conference on Information and Communication
  Technologies and Development}.\hskip 1em plus 0.5em minus 0.4em\relax ACM,
  2015, p.~43.

\bibitem{sharma2016prediction}
P.~Sharma and T.-S. Moh, ``Prediction of indian election using sentiment
  analysis on hindi twitter,'' in \emph{Big Data (Big Data), 2016 IEEE
  International Conference on}.\hskip 1em plus 0.5em minus 0.4em\relax IEEE,
  2016, pp. 1966--1971.

\bibitem{lu2014india}
D.~Lu, A.~Shah, and A.~Kulshrestha, ``India's\# twitterelection 2014​,''
  \emph{CS 224W Fall}, 2014.

\bibitem{gilani2016stweeler}
Z.~Gilani, L.~Wang, J.~Crowcroft, M.~Almeida, and R.~Farahbakhsh, ``Stweeler: A
  framework for twitter bot analysis,'' in \emph{Proceedings of the 25th
  International Conference Companion on World Wide Web}.\hskip 1em plus 0.5em
  minus 0.4em\relax International World Wide Web Conferences Steering
  Committee, 2016, pp. 37--38.

\bibitem{bhuwan-political-inclination}
\BIBentryALTinterwordspacing
B.~KC, ``Political inclination of journalits and it's influence on news,''
  Master's thesis, Ateneo de Manila University, the Philippines, 2014.
  [Online]. Available:
  \url{research.butmedia.org/?smd_process_download=1&download_id=318}
\BIBentrySTDinterwordspacing

\bibitem{donsbach2004political}
W.~Donsbach and T.~E. Patterson, ``Political news journalists,''
  \emph{Comparing political communication: Theories, cases, and challenges},
  pp. 251--270, 2004.

\bibitem{zaller1999theory}
J.~R. Zaller, ``A theory of media politics: How the interests of politicians,
  journalists, and citizens shapes the news,'' \emph{University of California,
  Los Angeles. Typescript}, 1999.

\bibitem{ali2018measuring}
A.~E. Ali, T.~C. Stratmann, S.~Park, J.~Sch{\"o}ning, W.~Heuten, and S.~C.
  Boll, ``Measuring, understanding, and classifying news media sympathy on
  twitter after crisis events,'' \emph{arXiv preprint arXiv:1801.05802}, 2018.

\bibitem{cha2010measuring}
M.~Cha, H.~Haddadi, F.~Benevenuto, and P.~K. Gummadi, ``Measuring user
  influence in twitter: The million follower fallacy.'' \emph{Icwsm}, vol.~10,
  no. 10-17, p.~30, 2010.

\bibitem{rao2015klout}
A.~Rao, N.~Spasojevic, Z.~Li, and T.~DSouza, ``Klout score: Measuring influence
  across multiple social networks,'' in \emph{Big Data (Big Data), 2015 IEEE
  International Conference on}.\hskip 1em plus 0.5em minus 0.4em\relax IEEE,
  2015, pp. 2282--2289.

\bibitem{manushree2017comparative}
A.~Manushree, M.~Adarsh, and P.~R. Kumar, ``A comparative method for different
  aspect based products features in online reviews of different languages,'' in
  \emph{Recent Trends in Electronics, Information \& Communication Technology
  (RTEICT), 2017 2nd IEEE International Conference on}.\hskip 1em plus 0.5em
  minus 0.4em\relax IEEE, 2017, pp. 1836--1840.

\bibitem{iguider2017language}
W.~Iguider and D.~R. Recupero, ``Language independent sentiment analysis of the
  shukran social network using apache spark,'' in \emph{Semantic Web Evaluation
  Challenge}.\hskip 1em plus 0.5em minus 0.4em\relax Springer, 2017, pp.
  129--132.

\bibitem{blei2003latent}
D.~M. Blei, A.~Y. Ng, and M.~I. Jordan, ``Latent dirichlet allocation,''
  \emph{Journal of machine Learning research}, vol.~3, no. Jan, pp. 993--1022,
  2003.

\bibitem{us1979belmont}
U.~D. of~Health, H.~Services \emph{et~al.}, ``The belmont report,'' 1979.

\bibitem{dittrich2012menlo}
D.~Dittrich, E.~Kenneally \emph{et~al.}, ``The menlo report: Ethical principles
  guiding information and communication technology research,'' \emph{US
  Department of Homeland Security}, 2012.

\bibitem{salganik2017bit}
M.~J. Salganik, \emph{Bit by bit: social research in the digital age}.\hskip
  1em plus 0.5em minus 0.4em\relax Princeton University Press, 2017.

\bibitem{le2017revisiting}
H.~T. Le, G.~Boynton, Y.~Mejova, Z.~Shafiq, and P.~Srinivasan, ``Revisiting the
  american voter on twitter,'' in \emph{Proceedings of the 2017 CHI Conference
  on Human Factors in Computing Systems}.\hskip 1em plus 0.5em minus
  0.4em\relax ACM, 2017, pp. 4507--4519.

\end{thebibliography}

\end{document}